\documentclass[11pt]{article}
\usepackage{amsmath,amssymb,color,graphics,epsfig,cite}
\usepackage{amsfonts}
\usepackage{xcolor}
\usepackage{multirow}

\newcommand{\hoch}[1]{$\, ^{#1}$}
\newcommand{\be}{\begin{equation}}
\newcommand{\ee}{\end{equation}}
\newcommand{\bea}{\begin{eqnarray}}
\newcommand{\eea}{\end{eqnarray}}

\begin{document}

\begin{center}
{\Large {\bf Observational Feasibility of 4D  Einstein-Gauss-Bonnet Cosmology: Bouncing and Non-Bouncing Universes}}
\vspace{20pt}

{\large H. khodabakhshi\hoch{1}, M. Farhang\hoch{2*} and H. L\"u\hoch{1,3}}

\vspace{10pt}

{\it \hoch{1}Center for Joint Quantum Studies and Department of Physics,\\ School of Science, Tianjin University,\\ Yaguan Road 135, Jinnan District, Tianjin 300350, China}

\bigskip
{\it \hoch{2} Department of Physics, Shahid Beheshti University, 1983969411, Tehran, Iran}

\bigskip
{\it \hoch{3}Joint School of National University of Singapore and Tianjin University,\\
International Campus of Tianjin University, Binhai New City, Fuzhou 350207, China}

\vspace{40pt}

\underline{ABSTRACT}
\end{center}

This paper analyzes the possibility of bouncing and non-bouncing universes in the framework of  four-dimensional Einstein-Gauss-Bonnet (4D-EGB) gravity,
corresponding respectively to negative and positive coupling constants $\lambda$ of the Gauss-Bonnet term. We also use the Horndeski-type scalar-tensor theory to assess the role of a scalar charge $C$ as a geometrical contribution to the radiation in the Universe.
We modify the expansion history of the universe to allow for modifications induced by the 4D-EGB gravity.
Using Planck measurements of the cosmic microwave background anisotropies as well as various datasets of baryonic acoustic oscillations, we set the upper bounds $\lambda \le 10^{-16} \text{(km/s/Mpc)}^{-2} $ and $\lambda \le 10^{-30} \text{(km/s/Mpc)}^{-2} $ for the non-bouncing and bouncing scenarios.
The upper limit in the latter case is mainly driven by the requirement to conservatively respect the thermal history at energy scales of the standard model of particle physics.
We also find that the contribution of the geometrical radiation-like term of the model cannot exceed 10\% of the current radiation in the Universe. The possibility of an early inflationary phase produced by a single scalar field is also studied and found to be feasible in both bouncing and non-bouncing scenarios.
This study shows the feasibility of a bouncing universe, even with a normal matter sector, in the 4D-EGB gravity.  More theoretical investigation is required to further explore possible observational predictions of the model that can distinguish between general relativity and 4D-EGB gravity.

\vfill{\footnotesize h\_khodabakhshi@tju.edu.cn \ \ \ m\_farhang@sbu.ac.ir (*Corresponding author)\ \ \ mrhonglu@gmail.com}

\thispagestyle{empty}
\pagebreak
\tableofcontents
\addtocontents{toc}{\protect\setcounter{tocdepth}{2}}

\section{Introduction}

General relativity (GR), a main pillar of the $\Lambda$CDM scenario, has passed many observational tests in a wide range of scales and plays a crucial role in understanding large-scale universe. However, given the conceptual challenges and observational tensions of $\Lambda$CDM, serious efforts have been put to exploring alternative gravities.
Among these is modification to the Einstein-Hilbert Lagrangian by integrating Gauss-Bonnet terms \cite{6}. These terms are important in low-energy gravity theories, especially in heterotic string theory \cite{7, 8} and studying quantum fields in curved spacetimes \cite{12}. A key feature of Gauss-Bonnet terms is their ability to produce ghost-free, second-order field equations \cite{6,13}.
Gauss-Bonnet terms do not alter gravitational field equations in 4-dimensional spacetime. However, there exists a  $D\rightarrow 4$ limit of Gauss-Bonnet gravity by rescaling the coupling constant \( \lambda \rightarrow \lambda/(D - 4) \) \cite{17}, as well as modifications through the Kaluza-Klein reduction \cite{14,18,19,15,20} that would change the  field equations in four dimensions. These approaches have led to novel insights into gravitational dynamics and important discoveries in cosmology \cite{20, 21, 22, 23, 24, 25, 26, 27, 28, 29}.

Several studies of 4-dimensional Einstein-Gauss-Bonnet (4D-EGB) gravity have led to various constraints on the parameter \( \lambda \). According to \cite{24}, inflation in the early Universe restricts negative values of \( \lambda \), with an upper limit suggested at \( 10^{-49} \, \text{(km/s/Mpc)}^{-2} \). On the other hand, \cite{25} notes that 4D-EGB gravity, with no cosmological constant, is challenged by current cosmological and gravitational-wave data. This study estimates \( \lambda \) to be between \( -10^{-20} \, \text{(km/s/Mpc)}^{-2} \) and \( 10^{-19} \, \text{(km/s/Mpc)}^{-2} \). In addition, employing the Affine Invariant Markov chain Monte Carlo (MCMC) Ensemble sampler, EMCEE \cite{27*}, and various datasets, \cite{27} constrains the upper bound of \( \lambda \) to be approximately \( 10^{-22} \, \text{(km/s/Mpc)}^{-2} \). Moreover, based on the need for 4D-EGB theory to be able to reproduce the lightest observed black holes, \cite{30} suggests that the maximum value of minus \(\lambda \) in 4D-EGB gravity should be near \( 10^{-28} \, \text{(km/s/Mpc)}^{-2} \). In these works, although positive and negative values for \(\lambda\) were considered, their cosmological consequences have not been explored.

In \cite{29} we explored the cosmological aspects of 4D-EGB gravity, minimally coupled to a perfect fluid, by considering both positive and negative values of \( \lambda \). It has been shown that selecting a negative value for \( \lambda \) results in a universe that begins with a bounce rather than the traditional Big Bang. This bouncing scenario notably does not violate the null energy condition, contrasting with many existing models. Furthermore, using the scalar-tensor reformulation and the Kaluza-Klein reduction \cite{14,15,17,19,20}, a scalar charge \( C \) is introduced \cite{18}, contributing a radiation-like component $\lambda C^4/H_0^2$ to the model. In \cite{rb}, using CMB data from the Atacama Cosmology Telescope (ACT) \cite{ACTPol}, it is shown that at the small-$\lambda$ limit, $\lambda C^4/H_0^2 = (-9 \pm 6) \times 10^{-6}$. This result leads to an imaginary field for positive values of $\lambda$. We will discuss how we can avoid an imaginary field in a bouncing universe.

In this paper, based on our results for the 4D-EGB model presented in \cite{29}, we aim to investigate observational tests for both bouncing and non-bouncing scenarios, examining the impact of the radiation-like component. To do so, we employ data from the temperature and polarization anisotropy spectra of the cosmic microwave background (CMB) radiation and its lensing, as measured by Planck \cite{31}, as well as the measurements of the Baryonic Acoustic Oscillations \cite{32}.
We also show that an early inflationary phase is conceivable for the bouncing and non-bouncing scenarios by including the scalar field's kinetic and potential terms into the 4D-EGB Lagrangian.

The structure of the paper is as follows: In Section~\ref{sec:model} we present the equations that describe the expansion history in the 4D-EGB model.
Section~\ref{sec:method} briefly introduces the methodology for data analysis as well as the datasets used in this work. The results are presented in  Section~\ref{sec:results} and we discuss the results and conclude in Section~\ref{sec:discussion}.
\section{Model}\label{sec:model}
\begin{figure}
	\centering
	\includegraphics[width=0.7\textwidth]{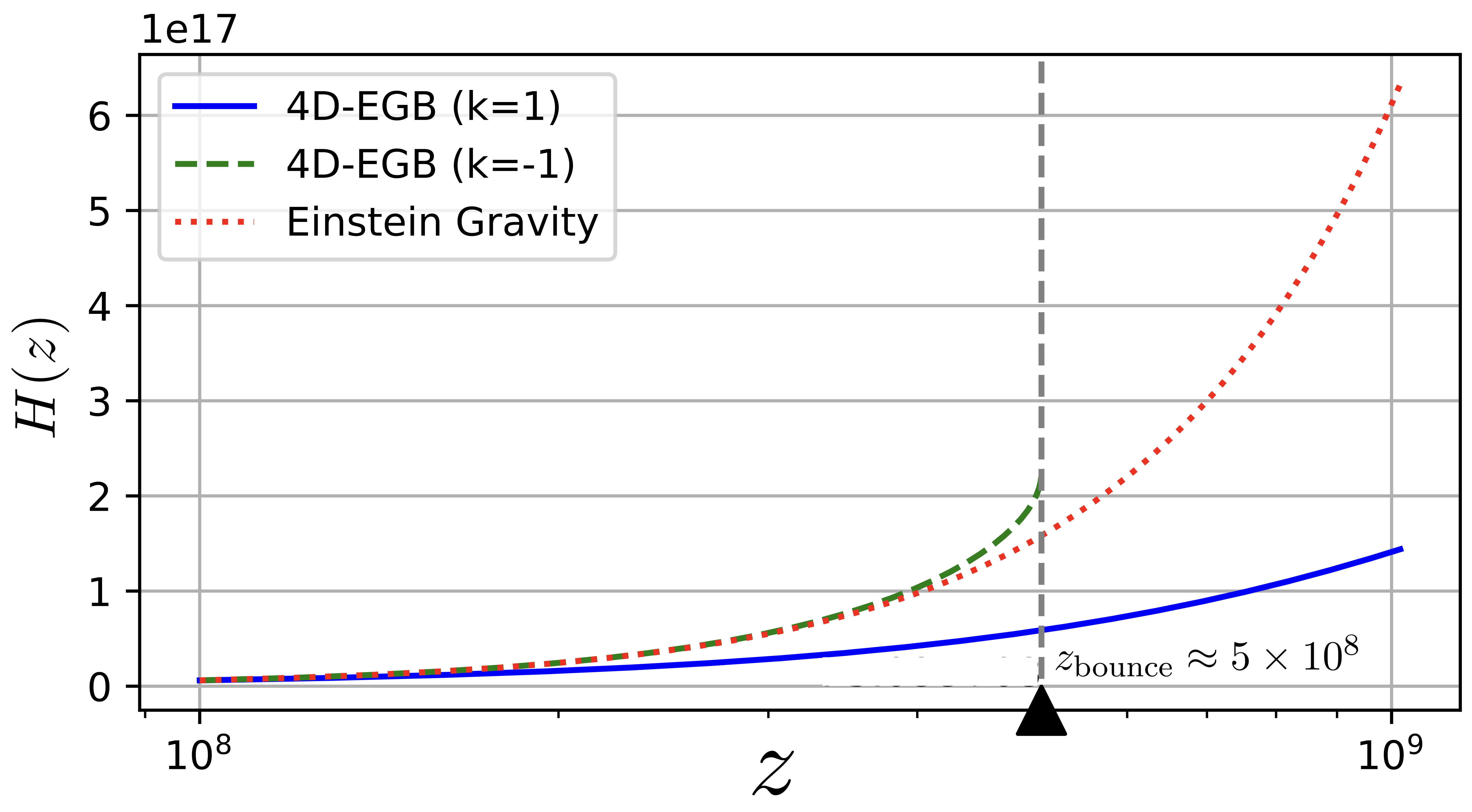}
	\caption{\small The Hubble parameter \( H \) as a function of redshift \( z \) in the very early universe  (\( 10^8 < z < 10^9 \)), plotted for Einstein gravity (red dotted-line) and the 4D-EGB gravity in a non-bouncing (blue solid-line) and  bouncing universe (green dashed-line). In this plot we consider \( \lambda = 10^{-35} \text{(km/s/Mpc)}^{-2} \) for both non-bouncing and bouncing universes. The bounce redshift can be obtained from eq. (\ref{rd}). Smaller \( \lambda \)'s would lead to  bounce at higher redshifts. The \(\Lambda\)CDM standard parameter values used in this plot are based on the bestfit measurements by Planck 2018 \cite{32}.}
	\label{fig:0}
\end{figure}

The 4D-EGB gravity action in the framework of the four-dimensional Horndeski-type scalar-tensor theory is described by \cite{14,15}
 \begin{equation} \label{action}
	\mathcal{A}=\int d^4 x \sqrt{-g}\big(R+k \lambda L_{\text{\tiny{4D-GB}}}\big),
\end{equation}
where 
 \begin{equation} \label{L4dgb}
L_{\text{\tiny{4D-GB}}}=\phi L_{\text{\tiny{GB}}} + 4 G^{ab} \nabla_a \phi \nabla_b \phi -4 \Box \phi (\nabla \phi)^2 +2 (\nabla \phi)^4
\end{equation}
\(k = \pm 1\), and \(\lambda \ge 0 \) represents the coupling constant which, depending on the choice of $k$, would lead to positive or negative coupling. $L_{\text{\tiny{GB}}}=R^2-4R_{ab}R^{ab}+R_{abcd}R^{abcd}$ is the Gauss-Bonnet term and $G^{ab}$ is the Einstein tensor and the curvature of the internal space is considered zero \cite{ 17, 18, 19, 20}. Considering a minimally-coupled perfect fluid matter with energy-momentum tensor $T^a_b = \text{diag}(\rho, p, p, p)$, in the context of FLRW metric, the time-time component of the Einstein field equations can be expressed as \cite{17, 18, 29}
\begin{equation}\label{eq:1}
	\left(\frac{\dot{a}}{a}\right)^2 + k \lambda \left[\left(\frac{\dot{a}}{a}\right)^4 - \frac{C^4}{a^4}\right]= \frac{8 \pi G}{3} \rho+  \frac{\Lambda c^2}{3}.
\end{equation}
The constant \(C\) is the scalar charge associated with the shift symmetry $\phi\rightarrow \phi +$const in the framework of four-dimensional Horndeski-type scalar-tensor theory. The cosmological constant is denoted as \(\Lambda\), and \(\rho = \rho_{\rm m} + \rho_{\rm r}\) is the density of dark and baryonic matter and radiation in our universe. The Friedmann equation (\ref{eq:1}) can be reformulated as
\begin{equation} \label{eq:2}
	H^2 + k\lambda H^4 = \frac{\Lambda c^2}{3} + \frac{8 \pi G}{3}(\rho_{\rm m} + \rho_{\rm r})  + \frac{k \lambda C^4}{a^4} \, .
\end{equation}
The eq. (\ref{eq:2}) yields two positive solutions for \(H^2\),
\begin{align}\label{ee1}
	H^2(z, \lambda) =  -\frac{1 + \epsilon \sqrt{1 + 4 k \lambda \left[ H_{{\text{ \tiny{E}}}}^2 (z) + k \lambda C^4 (1 + z)^4 \right]}}{2 k \lambda} , \hspace{1cm} \epsilon = \pm1,
\end{align}
where the Einstein Hubble \( H_{{\text{\tiny{E}}}} \) is defined as
\begin{equation} \label{eq:He2}
	H_{{\text{\tiny{E}}}}^2 = H_0^2 \left[ \Omega_{\rm r} (1 + z)^4 + \Omega_{\rm m} (1 + z)^3 + \Omega_\Lambda \right],
\end{equation}
and \(H_0\) is the Hubble constant and \(z = -1 + 1/a \) represents redshift.
The density parameters for $\Lambda$, matter and radiation are defined as
\begin{equation}
 \Omega_\Lambda = \frac{c^2 \Lambda}{3 H_0^2}\,, \hspace{1cm} \Omega_{\rm m}= \frac{8 \pi G}{3 H_0^2} \rho_{0{\rm m}}\,, \hspace{1cm} \Omega_{\rm r}= \frac{8 \pi G}{3 H_0^2} \rho_{0{\rm r}}\,,
 \end{equation}
 where $\rho_{0{\rm m}}$ and $\rho_{0{\rm r}}$ are the current density of matter and radiation.
 In eq.~(\ref{ee1}), the radiation-like term with the coefficient $\lambda C^4$ can be absorbed into the radiation term in $H_{\text{E}}^2$ through the redefinition
 \begin{equation}\label{rad}
 	\Omega'_\text{r} =  \Omega_\text{r} + k \frac{\lambda C^4}{H_0^2},
 \end{equation}
 where scalar charge $C$ is treated as a free parameter and is allowed to take both positive and negative values. For simplicity, in this section we will consider $C=0$. However, in the rest of the paper and for comparison with data, we will aslo consider $C \neq 0$ case.
 The universe would experience very different histories depending on the choice of $k$.
 For a comprehensive explanation of these solutions and their properties, including how they impact the dynamics of the universe, a detailed study is provided in \cite{29}. Here we briefly review some of the results.
 The solution with $\epsilon=1$ does not converge to Einstein limit when the dimensionless parameter $\lambda H_{{\text{\tiny{E}}}}^2$ goes to zero. We therefor focus only on scenarios with $\epsilon=-1$. The  \( k = 1 \) case corresponds to a universe starting from a singular point, the so called Big Bang, and expanding over time.
 On the other hand, the solution for \( k = -1 \) represents a bouncing universe, suggesting that the universe avoids the initial  singularity by going through a bounce. The bounce is shown to happen when \( H^2(z_0) = 1/(2\lambda) \), or, equivalently when
\begin{equation}\label{rd}
	1 - 4 \lambda H_{{\text{\tiny{E}}}}^2 (z_0) = 0,
\end{equation}
where $z=z_0$ is the redshift of the bouncing point (see eq.~(\ref{ee1})).
Using this equation, the value of \( \lambda \) can be determined if the bounce redshift is given. This point is illustrated in Fig. \ref{fig:0}.

The Hubble parameter as given by eq.~(\ref{ee1}), with $\epsilon=-1$, would converge to the Einstein limit as \(\lambda  H_{{\text{\tiny{E}}}}^2\) approaches zero
\begin{equation}
	H^2=H_{{\text{\tiny{E}}}}^2 - k \lambda (H_{{\text{\tiny{E}}}}^2)^2 + \ldots \, .
\end{equation}
The difference between this new \( H \) for both $k=\pm1$ and \( H_{{\text{\tiny{E}}}} \) would increase as one goes to  higher redshifts (see Fig. \ref{fig:0}).

\begin{figure}
	\centering
	\includegraphics[width=0.7\textwidth]{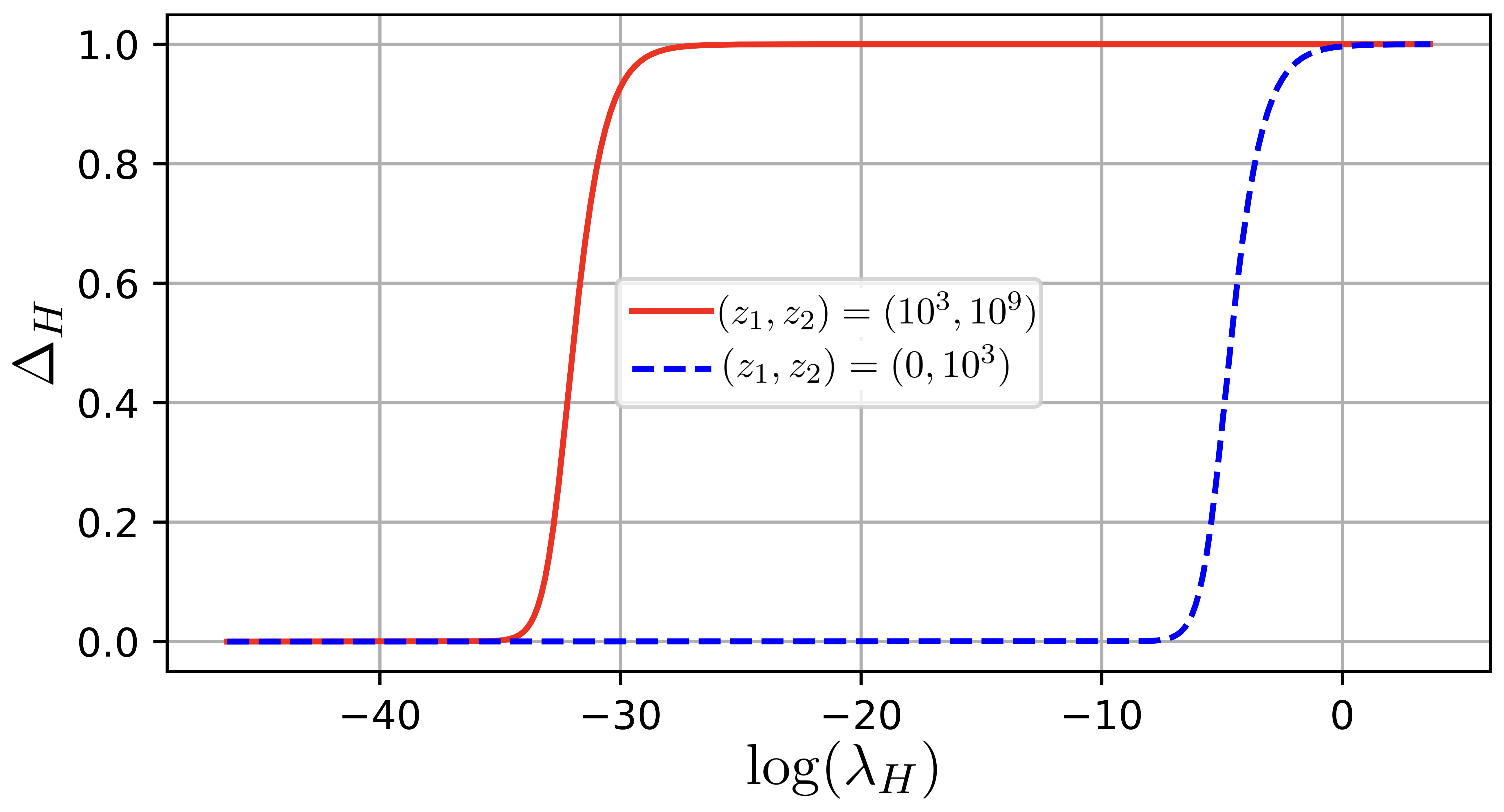}
	\caption{\small The maximum relative deviation $\Delta_H$ for \( z_1 = 10^3 \), \( z_2 = 10^9 \) (red solid-line), and for \( z_1 = 0 \), \( z_2 = 10^3 \) (blue dashed-line) are plotted for \( k = 1 \) and various \( \log(\lambda_{H}) \) values. Here, \( \lambda_H = \lambda H_0^2 \) parameterizes the dimensionless coupling constant. The deviation from the \(\Lambda\)CDM model is significant for \( \lambda \approx 10^{-10} \text{(km/s/Mpc)}^{-2} \) at \( z_2=10^3 \), while considering higher redshifts at \( z_2=10^9 \) constrains the upper limit of \( \lambda \) to approximately \( 10^{-37} \text{(km/s/Mpc)}^{-2} \). The \(\Lambda\)CDM standard parameter values used in this plot correspond to the bestfit measurements by Planck 2018 \cite{32}.}
	\label{fig:1}
\end{figure}

Given the observational achievements of the Einstein gravity, one may require that the deviation of 4D-EGB gravity from Einstein gravity stays  small even in the early universe. This can be ensured by choosing small enough values for \( \lambda \). In Fig. \ref{fig:1} for the case \( k=1 \), we illustrate the maximum relative deviation of 4D-EGB gravity from Einstein gravity as
\begin{equation}\label{eq3}
	\Delta_H(\lambda;z_1, z_2) = {\rm max}\bigg(\left| \frac{H^2(z, \lambda) - H_{{\text{\tiny{E}}}}^2(z)}{H_{{\text{\tiny{E}}}}^2(z)} \right|\bigg)\bigg|_{z_1}^{z_2},
\end{equation}
where the max is taken over $z$ in an assumed redshift range $z_1$ to $z_2$. The dimensionless parameter $\lambda_H$ in Fig. \ref{fig:1} is defined as $\lambda_H \equiv \lambda H_0^2$.
 The figure shows the high sensitivity of the 4D-EGB model to the choice of $\lambda$. The root of this sharp sensitivity to lambda is related to the different behavior of \( H^2/(1+z)^4 \rightarrow 0 \) and \( H_{{\text{\tiny{E}}}}^2/(1+z)^4 \rightarrow H_0^2 \Omega_\text{r} \) as \( z \rightarrow \infty \).

The overall behavior of $\Delta_H$ for the $k=-1$ case is quite similar to that of the $k=1$ case, but with different values for $\lambda$, which would in turn determine the redshift of the bouncing point in the early universe, as shown in eq. (\ref{rd}).
Caution should be taken when choosing $\lambda$ if the new scenario is required not to violate the thermal history of the Universe. In this work we choose to respect the thermal history and therefore assume that the bounce, if any, has occurred before, $z_{\rm bounce} \gtrsim 10^{10} $.
 In the following sections we investigate observational constraints on $\lambda$ for both the bouncing and non-bouncing universes.
 We also explore the implications for the scalar field $\phi$, or equivalently for the $\lambda C^4$ term —as a correction to radiation.
\section{Analysis and Datasets}\label{sec:method}

In our implementation of the 4D-EGB model, we modified the public Cosmological Monte Carlo (CosmoMc) package, originally designed for the $\Lambda$CDM model, to include the expansion history in the 4D-EGB scenario. There are two new parameters in this model: the coupling constant parameterized by $\lambda_{H}$ and the new relativistic degree of freedom that characterizes the scalar charge (see eq.~(\ref{rad})).
We perform the search for $\lambda_{H}$ logarithmically to cover an exhaustively broad range in the parameter space.
One should note that for the bouncing universe, corresponding to the $k=-1$ case, \(H^2\) becomes imaginary at high redshifts if \(\lambda\) is not sufficiently small. We discard these non-physical $\lambda$'s in the analysis by associating to them an {\it a priori} zero probability.
The highest allowed value for $\lambda$ in each scenario would also set the lowest possible redshift for the bounce.
 We request this bounce to happen before the highest redshift used in the analysis, i.e., $z <10^{10}$, which would correspond to $\lambda_H \lesssim 10^{-30}$, If the values for the other parameters are set based on the bestfits from Planck 2018 \cite{32}.
The degree of freedom corresponding to the scalar charge is included in the analysis by letting the effective number of relativistic species, \( N_{\text{eff}} \), be a free parameter. Changing \( N_{\text{eff}} \) would effectively alter the radiation density as
\begin{equation}
	\Omega_{\text{r}}= \left( 1+ \frac{7}{8} \left(\frac{4}{11}\right)^{4/3} N_{\text{eff}} \right) \Omega_{\gamma}, \hspace{1.5cm}   \Omega_{\gamma}= \frac{8 \pi G}{3 H_0^2} \rho_{\gamma} ,
\end{equation}
where  $\rho_{\gamma} $ is the energy density of photons.
The  fiducial value for the effective number of relativistic species, \( N_{\text{eff}}^{\rm fid} = 3.046 \),  corresponds to the standard scenario with  three flavors of massless neutrinos.
Thus, in this parametrization, any deviation  $\Delta N_{\rm eff}$ from the fiducial value would imply non-zero contribution of the radiation-like term \( \lambda C^4 \) to the radiation density,
\begin{equation}\label{lambdac}
	\frac{\lambda C^4}{H_0^2} \approx 0.2\,  \Omega_{\gamma}\, |\Delta N_{\rm eff}|.
\end{equation}

To constrain the parameters of the 4D-EGB model along with the standard cosmological parameters,  the predictions of the model need to be compared with observations.
As the departure from Einstein gravity grows with redshift, we expect the tightest constraint to come from the early universe data.
We therefor use the temperature and polarization power spectra of CMB anisotropies (TT+TE+EE) and CMB lensing as measured by Planck 2018 \cite{32}, as well as various measurements of the baryonic acoustic oscillations (BAO). We note that we carried two sets of parameter estimation, with CMB data alone, and with the combined CMB+BAO.  Since the implications for the 4D-EGB model were found to be consistent and indeed quite similar, we only report the results with the combined data. We refer to this combined dataset as P18+BAO.

\section{Results} \label{sec:results}
\begin{figure}
 	\centering
 	\includegraphics[width=1\textwidth]{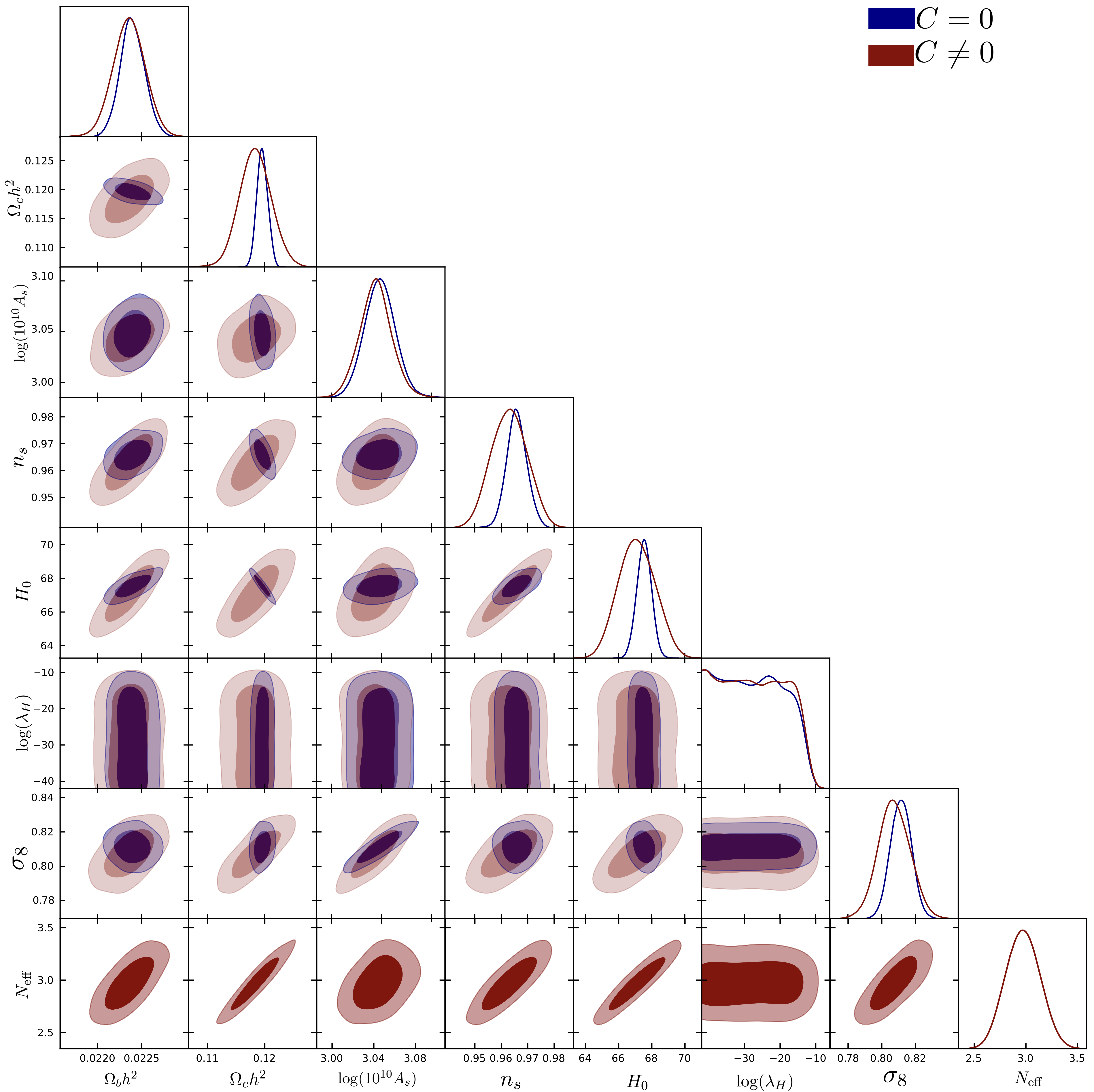}
 	\caption{\small Marginalized $1$D and $2$D likelihoods of the parameters in the 4D-EGB cosmology, as measured by P18+BAO,  for a non-bouncing universe ($k=1$), with $C=0$ and a free $C$, parametrized through $N_{\text{eff}}$ due to the radiation-like behavior of this degree of freedom.}
 	\label{fig:2}
 \end{figure}

In this section we use observations of CMB anisotropies and BAO to constrain the  parameters in the cosmological scenario based on the 4D-EGB gravity.
 We separately explore two cases of non-bouncing ($k=1$) and bouncing ($k=-1$) universes. In both scenarios, the
 parameter $\lambda$ characterizes the deviation from GR, and we also investigate the impact of the scalar charge by comparing the results for  $C=0$ and a varying $C$ case.
 Table~\ref{T1} and Figures~\ref{fig:2} and~\ref{fig:3} summarize these results.
 The second column of the table presents the $\Lambda$CDM measurements of the cosmological parameters, as the baseline for comparison.
  It should be noted that any modification to $N_{\rm eff}$ in this work is mainly associated to the scalar charge, and is thus expected to be absent in the Einstein gravity. This should be contrasted to various extensions to $\Lambda$CDM where variation in $N_{\rm eff}$ is associated to non-standard radiation-like contribution to the energy-momentum tensor of the Universe.
 In the table, we report the measurements for both $\lambda$ and $\log(\lambda_H)$, while in the plots we only present $\log(\lambda_H)$. The former is more conveniently used for comparison with the quoted constraints in the literature while the latter is numerically easier to work with.

\begin{table}
	\centering
	\begin{tabular}{cccccc}
		\hline \hline
		& \multicolumn{1}{c}{$\Lambda$CDM} & \multicolumn{2}{c}{4D-EGB, $k=1$} & \multicolumn{2}{c}{4D-EGB, $k=-1$} \\
		& & \( C=0\) & \( C \neq 0\) & \( C=0\) & \( C \neq 0\) \\
		\hline
		\( \Omega_{\rm b} h^2\) & {\scriptsize \(0.02242\pm0.00014\)} & {\scriptsize \(0.02240\pm0.00028\)} & {\scriptsize \(0.02236\pm0.00035\)} & {\scriptsize \(0.02240\pm0.00026\)} & {\scriptsize \(0.02235\pm0.00035\)} \\
		\(\Omega_{\rm c} h^2\) & {\scriptsize \(0.11933\pm0.00091\)} & {\scriptsize \(0.11960\pm0.00185\)} & {\scriptsize \(0.11840\pm0.00560\)} & {\scriptsize \(0.11960\pm0.00180\)} & {\scriptsize \(0.11820\pm0.00540\)} \\
		\(H_0\) & {\scriptsize \(67.66\pm0.42\)} & {\scriptsize \(67.55\pm0.86\)} & {\scriptsize \(67.10\pm2.10\)} & {\scriptsize \(67.55\pm0.80\)} & {\scriptsize \(67.10\pm2.10\)} \\
		$\lambda$  & $-$ &  \scriptsize $<10^{-17}$ & \scriptsize $<10^{-16}$  & \scriptsize $<10^{-31}$  & \scriptsize $10^{-30}$  \\
		\(\tau\) & {\scriptsize \(0.0561\pm0.0071\)} & {\scriptsize \(0.0570\pm0.0150\)} & {\scriptsize \(0.0560\pm0.0150\)} & {\scriptsize \(0.0570\pm0.0135\)} & {\scriptsize \(0.0560\pm0.0135\)} \\
		\(N_{\text{eff}}\) & $-$ & $-$ & {\scriptsize \(2.97\pm0.315\)} & $-$ & {\scriptsize \(2.96\pm0.325\)} \\
		\(\!\!\!\ln(10^{10}A_s)\) & {\scriptsize \(3.047\pm0.014\)} & {\scriptsize \(3.049\pm0.029\)} & {\scriptsize \(3.044\pm0.031\)} & {\scriptsize \(3.049\pm0.027\)} & {\scriptsize \(3.044\pm0.030\)} \\
		\(n_{\rm s}\) & {\scriptsize \(0.9665\pm0.0038\)} & {\scriptsize \(0.9659\pm0.0074\)} & {\scriptsize \(0.9630\pm0.0130\)} & {\scriptsize \(0.9657\pm0.0067\)} & {\scriptsize \(0.9630\pm0.0130\)} \\
		\hline
		$\log(\lambda_H)$ & $-$ & {\scriptsize \(<-13\)} & {\scriptsize \(<-12\)} & {\scriptsize $<-27$} & { \scriptsize$ <-26$} \\
		\(\sigma_8\) & {\scriptsize \(0.810\pm0.006\)} & {\scriptsize \(0.811\pm0.012\)} & {\scriptsize \(0.807\pm0.019\)} & {\scriptsize \(0.812\pm0.012\)} & {\scriptsize \(0.807\pm0.018\)} \\
		\hline
	\end{tabular}
	\caption{\small Constraints on the parameters of the 4D-EGB gravity for non-bouncing  ($k=1$) and bouncing universes ($k=-1$) measured by P18+BAO. In both scenarios, two cases of $C=0$ and free $C$ are explored which respectively correspond to no scalar charge and with a scalar charge associated to the shift symmetry of the scalar field in the theory. This radiation-like degree of freedom is parametrized through $N_{\rm eff}$. The $\Lambda$CDM results are also presented for comparison.}
	\label{T1}
\end{table}
 \subsection{Non-bouncing universe ($k=1$)} \label{sec:k1}
 The $k=1$ columns in Table~\ref{T1} present the results of parameter estimation with P18+BAO of a non-bouncing universe with a 4D-EGB gravity without a scalar charge (third column), and with the possibility of a scalar charge with its amplitude set as a free parameter to be measured by data (forth column). The $\Lambda$CDM scenario would correspond to the $\lambda\rightarrow 0$ limit of the 4D-EGB model. The parameter $\log (\lambda_H)$ is found to be consistent with zero, with tight upper bounds, which slightly relaxes as the scalar charge degree of freedom is included in the analysis.
 The measured mean value for $N_{\rm eff}$ is lower than expected from the standard model of particle physics $(N_{\rm eff}=3.046$). This is however slightly tricky to analyze.
 Firstly, the significance of this deviation is statistically low. Moreover, the reported $N_{\rm eff}$  from P18+BAO in the $\Lambda$CDM scenario is also low, i.e.,  $N_{\rm eff}=2.99\pm 0.33$ \cite{31}, again with low significance, yet with tighter errors compared to our results, as expected. It is therefore not clear that this small deviation can be associated to the new degrees of freedom allowed in the 4D-EGB model. Based on these results and using eq.~(\ref{lambdac}) we can estimate the order of magnitude of the upper bound on $\lambda C^4/H_0^2$, i.e., the radiation-like contribution to $H^2$ from this scenario (see eq.~(\ref{eq:2})). We find $\lambda C^4/H_0^2 \lesssim 10^{-6}$ ($1\sigma$), where $|\Delta N_{\rm eff}|\sim 10^{-1}$.

 \subsection{Bouncing universe ($k=-1$)}
 The \( k=-1 \) columns in Table~\ref{T1} present the results of parameter estimation with P18+BAO for the bouncing universe within the 4D-EGB gravity, corresponding to the $C=0$ and varying $C$ cases.
 The parameter \(\lambda\) is tightly constrained in both cases, with $\lambda \le 10^{-31} \text{(km/s/Mpc)}^{-2}$ for \(C=0\) and $\lambda \le10^{-30} \text{(km/s/Mpc)}^{-2}$ when marginalized over $C$.
 These constraints are significantly tighter compared to those in the non-bouncing scenario. It should be noted that the bounce redshift can be calculated for a given $\lambda$ from eq.~(\ref{rd}).
 For example, if we require the bounce to happen before the lepton era, with \(z \sim10^{15} \), and assuming that the \(\Lambda\)CDM parameters are set to their Planck bestfits, we find \(\lambda \approx 10^{-60} (\text{km/s/Mpc})^{-2}\).
 In our work we require $z_{\rm bounce} \gtrsim 10^{10}$ to guarantee the bounce does not happen earlier than the highest redshift directly used in the analysis. This would correspond to $\lambda \lesssim 10^{-35} \text{(km/s/Mpc)}^{-2} $ if $\Lambda$CDM parameter values are set to their Planck bestfits. Marginalizing over these parameters would relax this bound to the reported values in Table~\ref{T1}.

\begin{figure}
	\centering
	\includegraphics[width=1\textwidth]{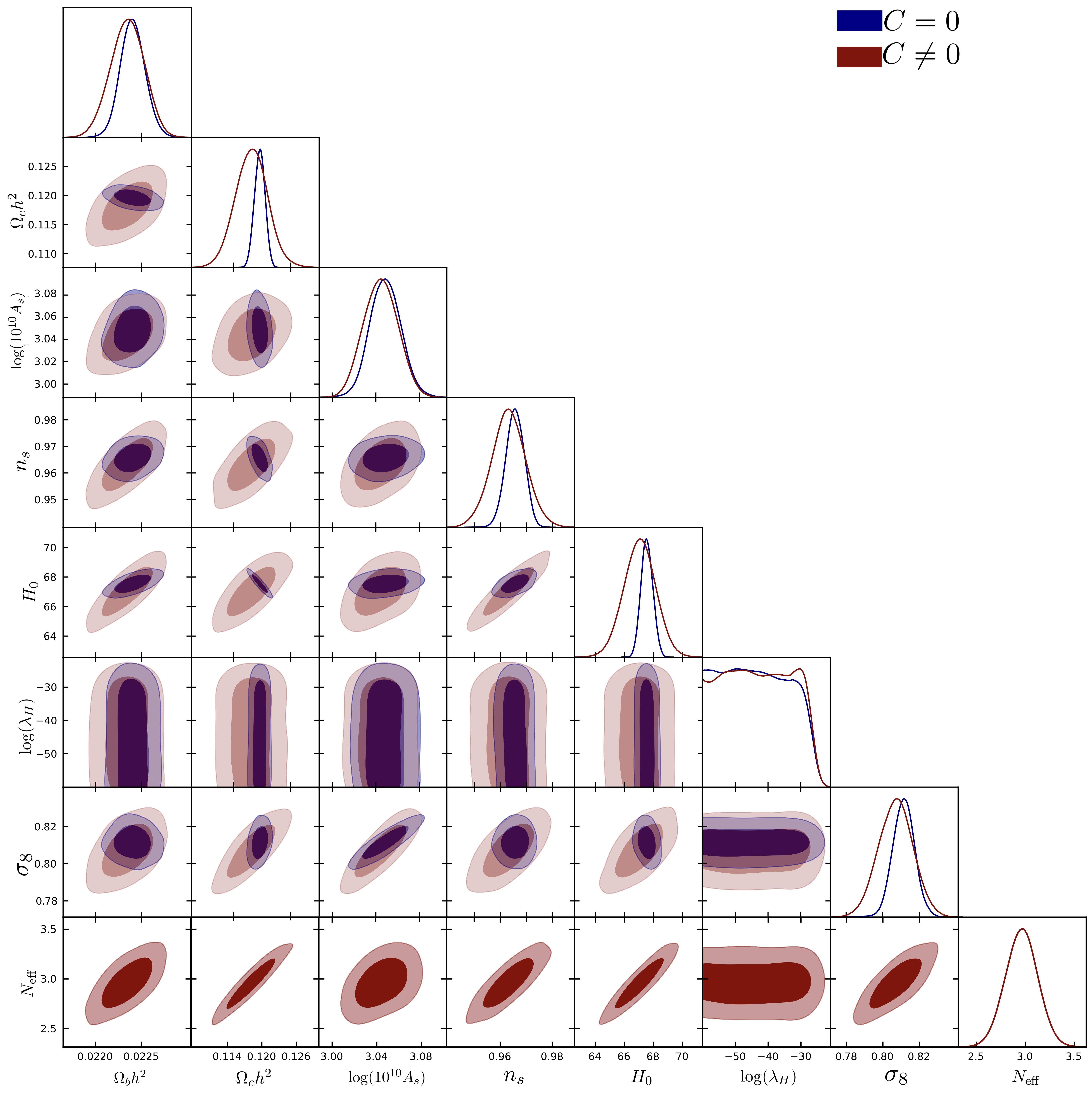}
	\caption{\small Similar to Figure~\ref{fig:2} but for  a bouncing universe ($k=-1$).}
	\label{fig:3}
\end{figure}
The order of magnitude of the upper bound on \(\lambda C^4/H_0^2\) for the $k=-1$ case is also similar to that found in Section~\ref{sec:k1}. It should however be noted that, unlike the $k=1$ scenario, the measured $N_{\rm eff}$ in this case should be lower than the fiducial $N_{\rm eff}$ to avoid an imaginary \(C\).
	
Unlike the Big Bang model, which starts from a singularity, a bouncing universe begins expansion from a non-singular state characterized by certain density and temperature which can be calculated from the bounce redshift.

Moreover, the early Universe is widely believed to have undergone an inflationary phase of rapid accelerated expansion. 
This phase is required to solve certain serious challenges faced by the standard hot big-bang model, 
the so-called horizon and flatness problems, and to seed cosmological inhomogeneities. We see that a radiation-filled Universe under the proposed 4D-EGB gravity does not undergo an inflationary phase in either bouncing or non-bouncing scenarios, as evidenced by $\ddot{a} < 0$. This is evident from the exact solution derived for $t(a)$ during the radiation-dominated era, calculated using eq.~(\ref{ee1}) \cite{29}. However, we show that an inflationary phase is indeed feasible in both bouncing and non-bouncing scenarios, if driven by a single canonical scalar field $\Phi$ with potential energy $V(\Phi)$ -similar to the inflationary scenario in a Universe with Einstein gravity (this is in contrast to the statement in \cite{25} that achieving inflation in the 4D-EGB model with $k=1$ is challenging).
 To show this point, we begin our investigation of inflation by adding both the kinetic term and the potential of the scalar field into the 4D-EGB Lagrangian as
\begin{equation} \label{inflation}
	\mathcal{A} = \int d^4 x \sqrt{-g} \left( R + k \lambda L_{\text{\tiny{4D-GB}}} - \frac{1}{2} \partial_a \Phi \partial^a \Phi - V(\Phi) \right).
\end{equation}
Variation of the above equation with respect to the FRW metric leads to the following Friedmann equations
\begin{equation} \label{eqf1}
	H^2 + k\lambda H^4 - \frac{k \lambda C^4}{a^4} = \frac{8 \pi G}{3} \left( \frac{1}{2} \dot{\Phi}^2 + V(\Phi) \right),
\end{equation}
\begin{equation} \label{eqf2}
	\dot{H} (1+2 k \lambda H^2) + \frac{2k \lambda C^4}{a^4} = -4\pi G \dot{\Phi}^2.
\end{equation}
The scalar field's equation of motion, obtained by varying the action with respect to $\Phi$ or combining eqs. (\ref{eqf1}) and (\ref{eqf2}), is
\begin{equation} \label{eqf3}
	\ddot{\Phi} + 3 H \dot{\Phi} + V_{,\Phi} = 0,
\end{equation}
where $V_{,\Phi} = dV/d\Phi$. Assuming $a(t)$ grows exponentially during inflation ($a(t) \approx e^{Ht}$), the terms $C^4/a^4$ become negligible in Friedmann eqs.~(\ref{eqf1}) and (\ref{eqf2}). 
Inflation  requires   $\epsilon \equiv -\dot{H}/H^2 \ll 1$ where $\epsilon$ is  the slow-roll parameter. During inflation, the potential energy of the field dominates over the kinetic energy, and the field evolves slowly, leading to the slow-roll approximations $\dot{\Phi}^2 \ll V(\Phi)$ and  $\ddot{\Phi} \ll 3 H \dot{\Phi}$. 
Applying these slow-roll conditions to eq. (\ref{eqf1}) and (\ref{eqf3}) yields
\begin{equation} \label{eq5}
	H^2 \approx \frac{8 \pi G}{3} \left( \frac{V(\Phi)}{1 + k\lambda H^2} \right),
\end{equation}
\begin{equation} \label{eq6}
	\dot{\Phi} \approx -\frac{V_{,\Phi}}{3H}.
\end{equation}
From eq. (\ref{eq5}), one can solve $H$ in terms of $V$ as
\begin{equation} \label{eq7}
	H^2 = \frac{-1 \pm \sqrt{1 + 4 k \lambda (8 \pi G V/3)}}{2 k \lambda},
\end{equation}
where the solution with the minus sign has no Einstein limit. 
In the search for solutions which are not drastically different from the Einstein case, we assume $Y \equiv \lambda (8 \pi G V / 3) \ll 1$ and 
expand eq.~(\ref{eq7}) to the second order in $Y$, as
\begin{equation} \label{eq8}
H^2 \approx \frac{8 \pi G V}{3} (1- k Y+  \mathcal{O}(Y^3) ).
\end{equation}
We will  see below that this is a reasonable assumption based on $\lambda$ values required for consistency with data.
For the case where $k = -1$, the bounce occurs at $H^2 = \frac{1}{2\lambda}$, corresponding to $Y = \frac{1}{4}$. This in turn determines $V_{\rm max} (\Phi)=V(\Phi_{\text{bounce}})$ for a chosen $\lambda$, also guaranteeing the condition $Y < \frac{1}{4}$ is met during inflation. Substituting eqs.~(\ref{eqf2}), (\ref{eq5}), and (\ref{eq6}) into the definition of the slow-roll parameter $\epsilon$, we obtain
\begin{equation} \label{eq9}
	\epsilon = \frac{1}{16\pi G} \left( \frac{V_{,\Phi}}{V} \right)^2 \frac{4 Y^2}{\sqrt{1-4kY} (-1+\sqrt{1-4kY})^2},
\end{equation}
which can be approximated to the second order in $Y$ for the case with the Einstein limit  as
\begin{equation} \label{eq10}
	\epsilon \approx \frac{1}{ 16 \pi G} \left( \frac{V_{,\Phi}}{V} \right)^2 (1 + Y^2 + \mathcal{O}(Y^3) ).
\end{equation}
This illustrates that the slow-roll parameter $\epsilon$ in 4D-EGB gravity corresponds to that in Einstein gravity up to the first order of $Y$. Therefore, as long as $Y \ll 1$ is satisfied (for small enough $\lambda$, determined by the choice of the inflationary potential),
 the condition $\epsilon < 1$ is also satisfied for various potential forms.

As a case study here we select the potential  $V(\Phi) = \frac{m^2 \Phi^2}{2}$. The number of e-folds during inflation is given by
$
N = \int_{\Phi_\text{f}}^{\Phi_\text{i}} \frac{H}{\dot{\Phi}} d\Phi.
$
Using eqs.~(\ref{eq6}) and (\ref{eq8}), we obtain
\begin{equation} \label{eq12}
	N \approx \frac{8 \pi G}{4} (\Phi_\text{i}^2 - \Phi_\text{f}^2) - k  \frac{(8 \pi G m)^2}{48} (\Phi_\text{i}^4 - \Phi_\text{f}^4) \lambda + \ldots,
\end{equation}
calculated up to the first order in $Y$. Assuming inflation progresses from $\epsilon = \epsilon_\text{i}$ to $\epsilon = \epsilon_\text{f} \approx 1$ allows us to determine $\Phi_\text{i}$ and $\Phi_\text{f}$ using eq.~(\ref{eq10}) as
\begin{equation} \label{eq13}
	\Phi_\text{f}^2 \approx \frac{1}{16 \pi G} , \quad \Phi_\text{i}^2 \approx \frac{1}{16 \pi G} \frac{1}{\epsilon_\text{i}}.
\end{equation}
Substituting the above into eq.~(\ref{eq12}) yields
\begin{equation} \label{eq14}
	N \approx \frac{1}{8} \left( \frac{1}{\epsilon_{\text{i}}} - 1 \right) -k \frac{m^2}{192} \left( \frac{1}{\epsilon_{\text{i}}^2} - 1 \right) \lambda + \ldots,
\end{equation}
where the first term corresponds to the Einstein case. Using $n_s \approx 0.9715$ from Table \ref{T1}, and taking  $N \approx \frac{2}{n_\text{s} - 1}$, we get $ N \approx 70$ which, using eq.~(\ref{eq14}), would give
\begin{equation} \label{eq15}
	\epsilon_{\text{i}} \approx \frac{12 + \sqrt{144 + k m^2 \lambda (-13462 + k m^2 \lambda)}}{13462 - k m^2 \lambda}.
\end{equation}
Considering the upper limits on $\lambda$ from Table \ref{T1}, and assuming $m \approx 10^{-6} (8\pi G)^{-1/2}$ \cite{last}, $\epsilon_\text{i} \approx 1.7 \times 10^{-3} \ll 1$ is obtained for both bouncing and non-bouncing scenarios. It should be noted that in the bouncing scenario, comparing the values for $\Phi_{\text{i}}$ and $\Phi_{\text{f}}$ with $\Phi_{\text{bounce}}$ indicates that inflation occurs after the bounce, and for both scenarios, the expansion condition $Y \ll 1$ is fully satisfied.

\section{Conclusions}\label{sec:discussion}
This study provided a comprehensive study of the observational tests of the 4D-EGB cosmology. It is shown \cite{29} that the 4D-EGB gravity model leads to two different scenarios: a bouncing and a non-bouncing universe, depending on the sign of the coupling constant (in our work represented by $k=1$ and $k=-1$ respectively). In the Horndeski-type scalar-tensor theory framework \cite{14,15,17,20}, the 4D-EGB model includes a scalar charge $C$, which contributes to the radiation density \cite{18}. We employed data from Planck 2018 (TT+TE+EE+lensing) and BAO \cite{31,32} to constrain both $\lambda$ and $\lambda C^4/H_0^2$ contribution to radiation and explored the cosmological aspects of the 4D-EGB model in its validation and predictions.
In this paper, we worked with the unit of $\lambda$ as $\text{(km/s/Mpc)}^{-2}$ and the unit of $C$ as equivalent to $\text{km/s/Mpc}$.

For a universe without a bounce, i.e. with a positive coupling constant, we find $\lambda \le 10^{-17} \text{(km/s/Mpc)}^{-2}$ with no scalar charge (i.e, with $C=0$) and  $\lambda \le 10^{-16} \text{(km/s/Mpc)}^{-2}$ if marginalized over $C$. The detailed results are listed in Table \ref{T1}  and illustrated in Figure \ref{fig:2}.
In a bouncing universe \cite{29}, where the initial singularity can be avoided, our analysis suggests tighter bounds of $\lambda \le 10^{-31} \text{(km/s/Mpc)}^{-2}$ and $\lambda \le 10^{-30} \text{(km/s/Mpc)}^{-2}$ if $C=0$ and if $C$ is marginalized over, respectively. See Table \ref{T1} and Figure \ref{fig:3}.
The bounds on \(\lambda\) in the latter case are mainly determined by the requirement of an early enough bounce, if it exists, occurring before the highest redshift used in our analysis (\(z_{\mathrm{bounce}} \gtrsim 10^{10}\)). This is a conservative choice, as a higher bounce redshift (and thus a lower \(\lambda\)) is required to respect the thermal history of the universe with an energy scale corresponding to the standard model of particle physics.
We also found that the contribution of the scalar charge to the radiation density is on the order of $10^{-1}$, indicating a significant influence in the overall radiation density of the universe.

In brief, the results are found to be consistent with general relativity, with no observational hints for deviations as suggested by a 4D-EGB gravity. However, the theoretical appeal of this theory, especially in the bouncing scenario where the initial singularity can be avoided, is encouraging for further investigation. In particular, the possibility of an early inflationary scenario in this model requires further exploration. We also note that the current work only searched for observational imprints of the model through its impact on the expansion history of the Universe and did not  investigate the modified perturbation equations \cite{rb}. The detailed study of these are left to future work.
We also investigated the possibility of an inflationary phase of early accelerated expansion  in the 4D-EGB model under the common assumption that inflation may be driven by a single canonical scalar field \(\Phi\) with potential energy \(V(\Phi)\).
We showed that achieving this phase is indeed possible  for both bouncing and non-bouncing scenarios given the small values of \(\lambda\) determined in our analysis.

The calculated bounds for \(\lambda\) are roughly consistent with previous works \cite{24,25,26,27,28,rb}. The feasibility of a bouncing universe within the framework of the 4D-EGB model provides new insights into the early Universe and demonstrates how modifications to gravity can alter our understanding of the Universe. Further theoretical investigation is required to fully understand the cosmological implications of the 4D-EGB model and the bouncing universe hypothesis.

\section*{Acknowledgement}

This work is supported in part by the National Natural Science Foundation of China (NSFC) grants No. 11935009 and No. 12375052

\end{document}